\documentclass[a4paper,11pt]{article}
\usepackage{pos}

\usepackage{natbib}
\title{
The Cherenkov Camera for the PBR mission }

\author*[a,b]{Valentina Scotti}
\author[b]{Antonio Anastasio}
\author[c,d]{Mario Bertaina}
\author[b]{Alfonso Boiano}
\author[e,f]{Rossella Caruso}
\author[g]{Cristian De Santis}
\author[b]{Vincenzo Masone} 
\author[a,b]{ Marco Mese}
\author[b]{Giuseppe Osteria}
\author[a,b]{Beatrice Panico}
\author[b]{Giuseppe Passeggio}
\author[a,b]{Francesco Perfetto}
\author[b]{Haroon Akhtar Qureshi}
\author[j,k]{Ester Ricci}
\onbehalf{for the JEM-EUSO Collaboration\\[-1mm]{\normalsize \normalfont (a complete list of authors can be found at the end of the proceedings)}}

\affiliation[a]{Dipartimento di Fisica "E. Pancini", Università di Napoli Federico II\\
}

\affiliation[b]{INFN, Sezione di Napoli\\
}

\affiliation[c]{INFN, Sezione di Torino\\
}

\affiliation[d]{Università degli studi di Torino\\
}

\affiliation[e]{INFN, Sezione di Catania\\
}

\affiliation[f]{Università degli studi di Catania\\
}

\affiliation[g]{INFN, Sezione di Roma Tor Vergata\\
}

\affiliation[j]{INFN, TIFPA\\
}

\affiliation[k]{Università degli studi di Trento\\
}

\emailAdd{scottiv@na.infn.it}

\abstract{The POEMMA-Balloon with Radio (PBR) mission is a NASA super-pressure balloon experiment designed to advance the detection of ultra-high-energy cosmic rays, high-altitude horizontal air showers, and astrophysical neutrinos. A key instrument of PBR is the Cherenkov Camera (CC), which utilizes a 2048-pixel SiPM camera to detect the optical Cherenkov emission from cosmic-ray-induced air showers and search for upward-going signals indicative of neutrinos.

The CC operates in the 320–900 nm spectral range with a 10 ns integration time, leveraging a bi-focal optical design to enhance detection efficiency. The CC enables precise reconstruction of shower trajectories and provides valuable data on cosmic rays' composition and energy distribution. PBR’s sub-orbital altitude is particularly advantageous for these measurements, offering a unique vantage point that bridges the observational gap between ground-based and space-based instruments. Additionally, the CC will play a critical role in neutrino searches, detecting tau-lepton decay showers from Earth-skimming neutrinos associated with astrophysical transients. By integrating the CC with fluorescence and radio detection systems, PBR will pioneer a multi-messenger approach to high-energy cosmic phenomena, refining observational techniques for future space-based missions.

This contribution will describe the current status of the development of the CC as well as its expected performance.}

\ConferenceLogo{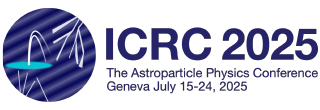}

\FullConference{39th International Cosmic Ray Conference (ICRC2025)\\
 15–24 July 2025\\
Geneva, Switzerland\\}

\begin{document}
\maketitle

\section{Introduction}

The detection of Ultra-High-Energy Cosmic Rays (UHECRs) and Very High Energy Neutrinos (VHENs) is a central challenge in astroparticle physics, offering a unique window into the most energetic processes in the universe. Recent efforts, summarized in the UHECR community white paper~\cite{coleman2023uhecrwp}, have identified the need for next-generation observatories capable of both high-precision measurements and unprecedented exposure at the highest energies. The Probe of Extreme Multi-Messenger Astrophysics (POEMMA)~\cite{poemma2021jcap}, a proposed dual-satellite mission, is designed to fulfill these requirements by observing extensive air showers (EASs) from orbit. As a vital step toward POEMMA, the POEMMA-Balloon with Radio (PBR) mission serves as a stratospheric pathfinder to validate key technologies and detection strategies in a near-space environment \cite{icrc2025}.

A core component of PBR is the Cherenkov Camera (CC), which is engineered to detect the optical Cherenkov emission from high-altitude horizontal air showers (HAHAs) and Earth-skimming tau neutrinos. The CC consists of 2048 silicon photomultiplier (SiPM) pixels arranged to conform to the telescope’s curved focal plane, enabling a wide field of view of 12° × 6° and high angular resolution (0.2° per pixel). The use of bi-focalizing optics reduces false triggers by requiring spatial coincidence, enhancing the reliability of Cherenkov event identification. The CC is complemented by a dual-polarized radio instrument (RI), together forming a multi-hybrid detection system capable of simultaneously capturing both optical and radio signatures of EASs~\cite{pueo2021jinst,cummings2021prd,cummings2021prd2}.

The PBR mission builds upon experience from previous balloon-borne efforts such as EUSO-SPB1~\cite{euso2024astropart} and EUSO-SPB2~\cite{euso2023icrc}, while incorporating several design upgrades including improved optics, broader wavelength sensitivity (320--900~nm), and advanced readout electronics under development. With this instrumentation, PBR aims to achieve three science goals: (1) to observe UHECRs via fluorescence emission from above, (2) to characterize HAHAs with combined Cherenkov and radio techniques, and (3) to search for tau neutrino signatures from astrophysical transient events such as gamma-ray bursts and neutron star mergers~\cite{fang2017apj,guepin2022nrp,venters2020prd}.

This paper focuses on the Cherenkov Camera subsystem, detailing its design, capabilities, and expected contributions to PBR’s mission objectives. By demonstrating Cherenkov detection of EASs from near-space, PBR will raise the technical readiness level for POEMMA and help establish the feasibility of a space-based multi-messenger observatory.

\section{Scientific Objectives of the Cherenkov Camera}

The Cherenkov Camera (CC) aboard the POEMMA-Balloon with Radio (PBR) mission is a critical instrument designed to perform pioneering measurements of extensive air showers (EASs) via detection of optical Cherenkov emission. Its primary scientific objectives focus on three main domains:

\begin{itemize}
    \item \textbf{Characterization of High-Altitude Horizontal Air Showers (HAHAs):} \\
    The CC will detect Cherenkov light from cosmic-ray-induced air showers developing entirely in the upper atmosphere (above 20~km), where reduced atmospheric density allows for extended propagation and minimal attenuation. These events are tightly beamed and can be observed with high efficiency using the PBR platform. The intensity and angular profile of the Cherenkov signal is sensitive to the primary cosmic ray composition, offering a means to statistically discriminate between light and heavy nuclei. Over a nominal month-long flight, PBR is expected to record up to 9,000 HAHA events, significantly advancing our understanding of EAS development at high altitudes and near the cosmic ray spectral knee ~\cite{proposal_main}.

  \item \textbf{Validation of Neutrino Detection Techniques:} \\
  The CC will search for upward-going EASs initiated by tau leptons generated from Earth-skimming tau neutrinos. These events, observable below the Earth’s limb, produce Cherenkov signatures analogous to HAHAs but arrive from below the horizon. Such measurements will validate detection strategies for space-based neutrino observatories and extend the sensitivity to very-high-energy neutrinos beyond the PeV scale. The CC’s capability to detect such events complements current ground-based efforts and enhances sensitivity to astrophysical neutrino point sources~\cite{cummings2021prd, cummings2021prd2}.

    \item \textbf{Participation in Multi-Messenger Transient Follow-Ups:} \\
    The CC will play an active role in rapid response to astrophysical alerts from facilities such as LIGO/Virgo/KAGRA, Fermi, and IceCube. By tilting the telescope’s field of view to follow event localizations near the Earth’s limb, PBR can search for optical Cherenkov signatures from neutrino-induced upward-going showers associated with transient phenomena including gamma-ray bursts, neutron star mergers, and blazar flares~\cite{venters2020prd, poemma2021jcap}.. These capabilities position the CC as a versatile instrument in the broader landscape of multi-messenger astrophysics.
\end{itemize}

\section{The Focal Surface}

\begin{figure}
    \centering
    \includegraphics[width=0.9\linewidth]{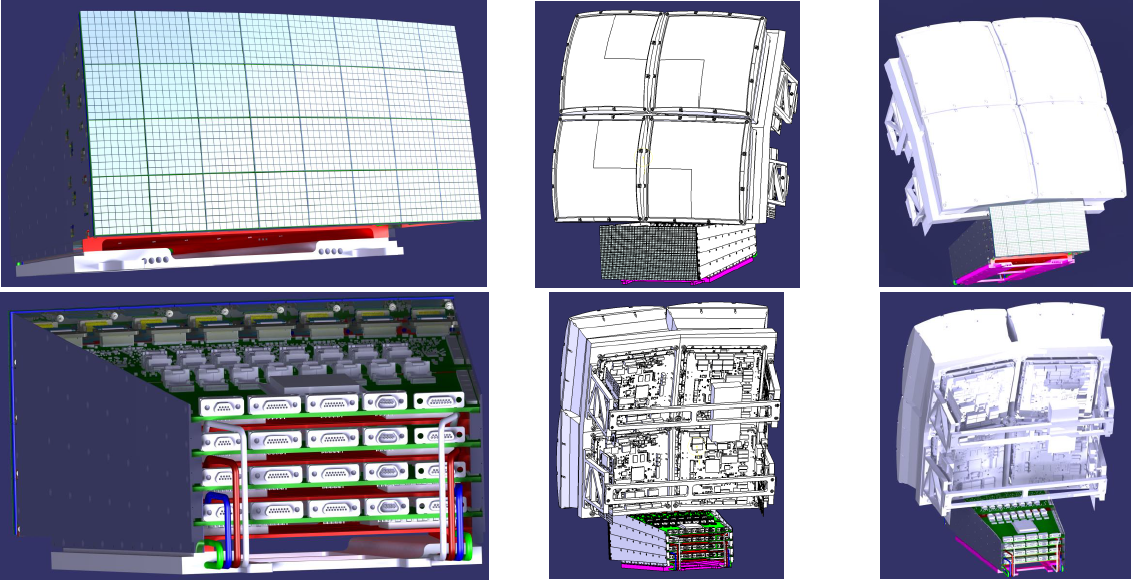}
    \caption{A cad view of the Focal Surface of PBR. Right panel: the Cherenkov camera and its electronics. Left panel: the Cherenkov Camera is positioned below the Fluorescence Camera. }
    \label{fig:CC_FS}
\end{figure}

The PBR Cherenkov Camera (CC) comprises four rows of 8 SiPM arrays . Each array consists of $8 \times 8$ channels of $3 \times 3~\mathrm{mm}^2$ pixels, for a total of 2048 pixels (see Figure \ref{fig:CC_FS}). The arrays are mounted on a structure that allows the detection plane to approximate the spherical focal surface required by the telescope optics. The camera has a field of view (FoV) of $12^\circ \times 6^\circ$, with each pixel subtending $0.2^\circ$. The current baseline for the $8 \times 8$ SiPM arrays is the Hamamatsu S13361-3050 model, sensitive to wavelengths in the range $320{-}900$~nm. 

\section{The Read-out Electronics}

The SiPM readout electronics are based on the MIZAR application-specific integrated circuit (ASIC) developed by INFN Turin. Each MIZAR chip supports 64 channels with an integrated front-end amplifier employing a common-gate topology, followed by 256 memory cells sampling at 200~MHz. Each cell contains a storage capacitor, a single-slope analog-to-digital converter with programmable resolution (7–12 bits), and digital control logic. The memory can be configured in groups of 32, 64, or 256 cells to optimize dead time. Each channel can be programmed with two thresholds; a trigger is generated if at least two pixels (as required by the bi-focal optical design) exceed the thresholds. The ASIC then stores the hitmap and sends a trigger request to a control FPGA. If accepted, the hit data are digitized and read out; otherwise, the ASIC resumes sampling. This trigger-on-validation design minimizes power usage. The ASIC is manufactured in 65~nm CMOS technology, with a power consumption of 10~mW per channel.

\begin{figure}
    \centering
    \includegraphics[width=0.9\linewidth]{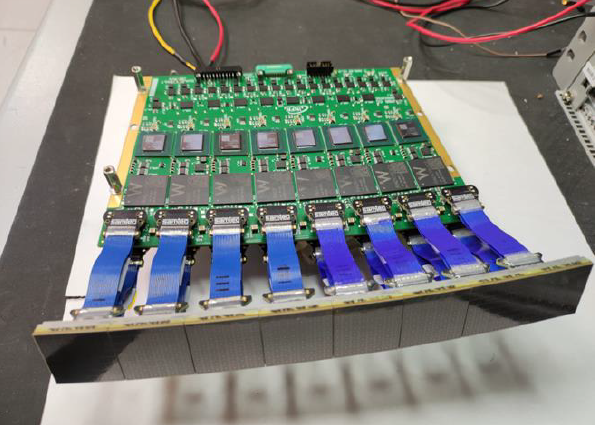}
    \caption{The Elementary Cell (CC-PDM) of the Focal Surface of the Cherenkov camera and its read-out electronics based on the Radioroc ASIC}
    \label{fig:CC_EC}
\end{figure}

Groups of eight ASICs along a row (a CC Photo Detection Module, or CC-PDM) are managed by a Xilinx FPGA that handles triggering, data packaging, and readout (see Figure \ref{fig:CC_EC}). The FPGA is mounted on a card that also contains a temperature-compensated bias voltage system for the SiPMs. A dedicated electronics board coordinates the trigger signals from the four FPGAs and provides synchronized timing across the entire camera.

At the moment, another design based on the Weeroc Radioroc ASIC is being studied at INFN Napoli. While this option does not allow to wavelength digitization, it offers a lower data budget solution. Both solutions are compatible with the same control FPGA board. The use of Radioroc ASIC allows for channel-by-channel adjustment of SiPM high-voltage, enabling fine SiPM gain adjustment to correct for channel gain non-uniformity. Additionally, this ASIC can trigger at levels as low as 1/3 photoelectron, offering dual-gain energy measurement capabilities, and maintains 1\% linearity in energy measurement up to 2000 p.e. 

All electronics are housed in a metallic rack behind the focal surface, which includes thermal regulation and shielding to minimize electromagnetic interference and to ensure stable operation in stratospheric conditions.

\section{Mechanics and optics}

The Cherenkov Camera is integrated into a Schmidt-type telescope with a 1.1~m aperture and a $\sim$1.6~m radius of curvature, optimized for operation at high altitudes. The optical system includes an aspheric corrector plate made of UV-transparent PMMA and a segmented primary mirror composed of 12 vacuum-slumped borosilicate glass elements with aluminum reflective coatings. The bi-focal optical design introduces an optical accordion element made from PMMA, consisting of a 1D prism array mounted 100~mm in front of the CC focal plane, which splits incoming light into two spatially separated spots to suppress background triggers. The point spread function is maintained below one pixel (3~mm) RMS across the expected incident angles, ensuring sharp imaging. 

The SiPM arrays of the CC are mounted to conform to the curved focal surface of the telescope using a precision mechanical support structure. In designing the Focal Surface mechanics, the goals were:
\begin{itemize}
    \item Minimize the dimensions of the Elementary Cell.
    \item Fix the ECs on a spherical-shaped metallic surface.
    \item Keep the connection between the EC and the electronics “flexible”.
\end{itemize}
The resulting mechanics allow for a good alignment of the SiPM arrays, while dead space among arrays is minimized.
allow
The full assembly is mounted to a lightweight, C-frame mechanical structure designed to allow elevation rotation from nadir to $12^\circ$ above the horizon, enabling flexible pointing for cosmic-ray and neutrino observations. 

\section{Currents status}

A prototype of the CC-PDM, complete with sensor and readout electronics, has been developed. A characterization campaign is currently underway to study the key performance parameters of the selected SiPM matrix, such as gain, crosstalk probability, photon detection efficiency (PDE) and afterpulse probability, in conjunction with the Cherenkov Camera (CC) electronics.

\begin{figure}
    \centering
    \includegraphics[width=0.8\linewidth]{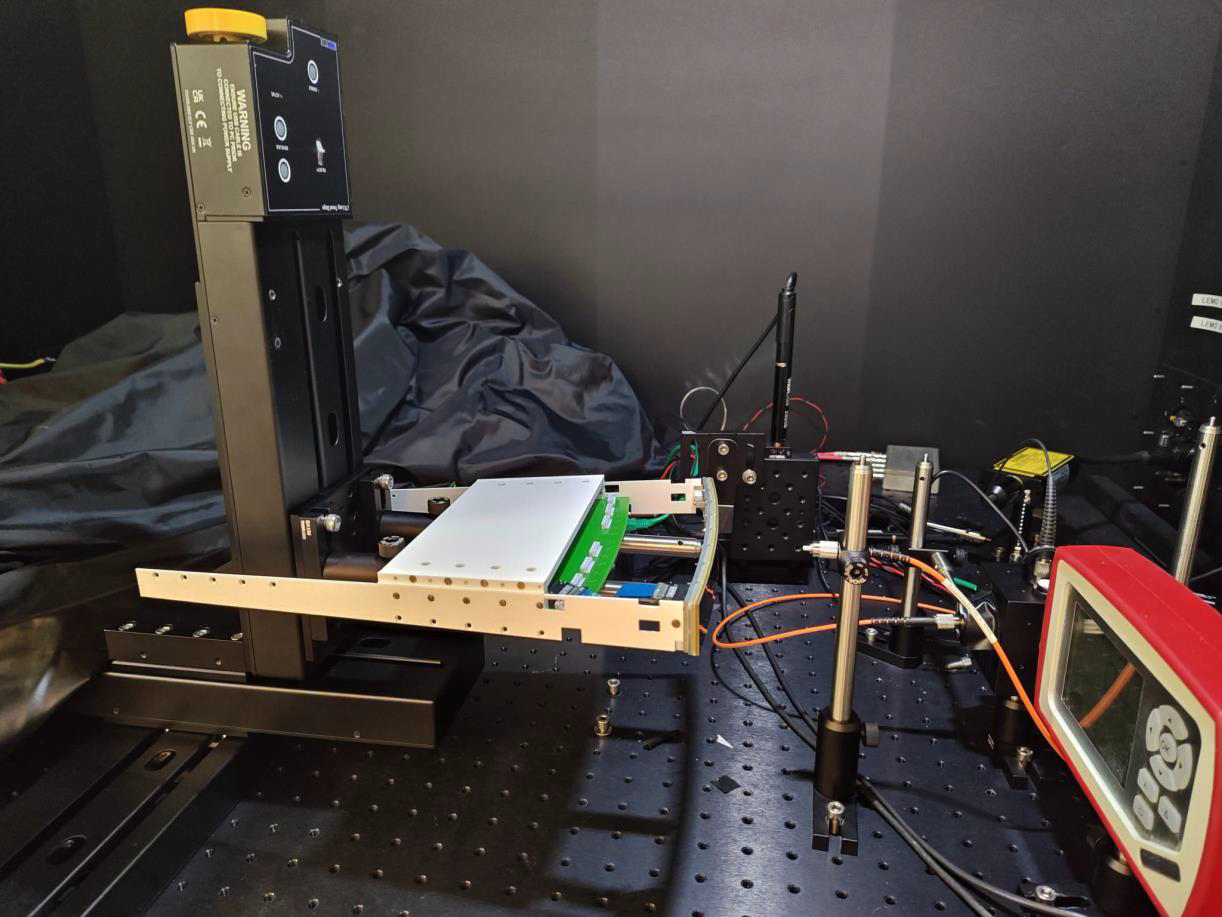}
    \caption{Experimental setup used to measure the PDE of the SiPM arrays. The optical fiber is connected to a Hamamatsu PLP-10 laser diode head (405~nm).}
    \label{fig:pde_meas}
\end{figure}

The experimental setup (Fig.~\ref{fig:pde_meas}) consists of an integrating sphere coupled to a laser source and a calibrated photodiode. A pinhole mounted on one of the ports in the sphere allows illumination of a single SiPM within the matrix. A remotely controlled 3-axis positioning system is used to scan all individual sensors. Measurements of the PDE for Hamamatsu arrays, performed using Radioroc readout electronics, are consistent with the specifications provided by the manufacturer.

In parallel, a comprehensive simulation effort is being conducted using the Geant4 framework, integrated into the offline analysis structure. CAD models of the mirrors and mechanical components have been incorporated into the simulation. The event generation is based on real EUSO-SPB2 data, while simulation studies are performed using the \texttt{EASCherSim*} package. Preliminary results indicate an energy threshold of approximately 0.5~PeV, with peak sensitivity near 3~PeV, and an energy-dependent angular acceptance. The number of expected events is around 65 per hour.

\section{Conclusions}

The PBR Cherenkov Camera represents a significant advancement in the development of stratospheric balloon-borne observatories. Designed to investigate high-altitude horizontal air showers and to detect Earth-skimming neutrinos in response to astrophysical event alerts, PBR builds on the heritage of POEMMA with its hybrid focal surface architecture. By combining fast optical response (10~ns), a wide field of view (12$^\circ$~$\times$~6$^\circ$), and a bi-focal optical design to suppress false triggers, the Cherenkov Camera aboard PBR will serve as both a pathfinder for future satellite missions and a groundbreaking observatory in its own right. Its operation will elevate the technical readiness of the POEMMA mission architecture and provide critical insights into the most energetic processes in the universe. This mission will not only contribute valuable data to multi-messenger astrophysics but also serve as a key step toward increasing the technical readiness of future space-based observatories. 

PBR is scheduled to launch as a NASA super pressure balloon payload from Wanaka, New Zealand, in the first half of 2027.

\section{Acknowledgements}
The authors acknowledge the support by NASA award 80NSSC22K1488 and 80NSSC24K1780, by the French space agency CNES and the Italian Space agency ASI. The technological development of the Cherenkov Camera was made possible through the ASI-INFN agreement n. 2021-8-HH.0 and its amendments 2021-8-HH.1-2021, n. 2021-8-HH.2-2022 and n. 2021-8-HH.3-2025. The work is supported by OP JAC financed by ESIF and the MEYS $CZ.02.01.01/00/22\_008/0004596$. This research used resources of the National Energy Research Scientific Computing Center (NERSC), a U.S. Department of Energy Office of Science User Facility operated under Contract No. DE-AC02-05CH11231. We also acknowledge the invaluable contributions of the administrative and technical staff at our home institutions.

    \newpage
{\Large\bf Full Authors list: The JEM-EUSO Collaboration}

\begin{sloppypar}
{\small \noindent
M.~Abdullahi$^{ep,er}$              
M.~Abrate$^{ek,el}$,                
J.H.~Adams Jr.$^{ld}$,              
D.~Allard$^{cb}$,                   
P.~Alldredge$^{ld}$,                
R.~Aloisio$^{ep,er}$,               
R.~Ammendola$^{ei}$,                
A.~Anastasio$^{ef}$,                
L.~Anchordoqui$^{le}$,              
V.~Andreoli$^{ek,el}$
% ...
}
\end{sloppypar}
\vspace*{.3cm}

{ \footnotesize
\noindent
%
% ...
$^{cb}$ Universit\'e de Paris, AstroParticule et Cosmologie (CNRS), Paris, France\\
% ...
$^{ef}$ Istituto Nazionale di Fisica Nucleare (INFN), Sezione di Napoli, Naples, Italy\\
% ...
$^{ei}$ Istituto Nazionale di Fisica Nucleare (INFN), Sezione di Roma Tor Vergata, Rome, Italy\\
% ...
$^{ek}$ Istituto Nazionale di Fisica Nucleare (INFN), Sezione di Torino, Turin, Italy\\
$^{el}$ Universit\`a di Torino, Dipartimento di Fisica, Turin, Italy\\
% ...
$^{ep}$ Gran Sasso Science Institute (GSSI), L’Aquila, Italy\\
$^{er}$ Istituto Nazionale di Fisica Nucleare (INFN), Laboratori Nazionali del Gran Sasso, Assergi, Italy\\
% ...
$^{ld}$ University of Alabama in Huntsville, Huntsville, AL, USA\\
$^{le}$ City University of New York (CUNY), Lehman College, Bronx, NY, USA\\
% ...
}

\end{document}